\documentclass[aps,pre,reprint,superscriptaddress]{revtex4-2}
\usepackage{amsmath}
\usepackage{graphicx}
\usepackage{dcolumn}
\usepackage{bm}
\usepackage{hyperref}
\usepackage{courier}
\usepackage{braket}
\usepackage{float}
\usepackage[usenames,dvipsnames]{xcolor}
\hypersetup{
  colorlinks   = true,	
  urlcolor     = blue,	
  linkcolor    = blue,	
  citecolor   = blue	
}
\newcommand{\cc}{\color{blue}}
\graphicspath{{fig2/}}
\begin{document}
\newcommand{\ve}{\varepsilon}
\title{Quantum Turing bifurcation: Transition from quantum amplitude death to quantum oscillation death}

\author{Biswabibek Bandyopadhyay}
\affiliation{Chaos and Complex Systems Research Laboratory, Department of Physics, University of Burdwan, Burdwan 713
  104, West Bengal, India}
\author{Taniya Khatun}
\affiliation{Chaos and Complex Systems Research Laboratory, Department of Physics, University of Burdwan, Burdwan 713
  104, West Bengal, India}
\author{Tanmoy Banerjee}
\email[]{tbanerjee@phys.buruniv.ac.in}
\altaffiliation{he/his/him}
\affiliation{Chaos and Complex Systems Research Laboratory, Department of Physics, University of Burdwan, Burdwan 713
  104, West Bengal, India}

\date{\today}

\begin{abstract}
An important transition from a homogeneous steady state to an inhomogeneous steady state via the Turing bifurcation in coupled oscillators was reported in [{\cc Phys. Rev. Lett. {\bf 111}, 024103 (2013)}]. However, the same in the quantum domain is yet to be observed. 
In this paper, we discover the quantum analogue of the Turing bifurcation in coupled quantum oscillators. We show that a homogeneous steady state is transformed into an inhomogeneous steady state through this bifurcation in coupled quantum van der Pol oscillators. We demonstrate our results by a direct simulation of the quantum master equation in the Lindblad form. We further support our observations through an analytical treatment of the noisy classical model. Our study explores the paradigmatic Turing bifurcation at the quantum-classical interface and opens up the door towards its broader understanding.   
\end{abstract}

\maketitle

\section{Introduction}
The Turing bifurcation \cite{turing} leads to heterogeneity from a homogeneous solution through a spontaneous symmetry breaking mechanism and has been widely studied in the context of pattern formation in chemistry, biology, and physics \cite{murry2,tur-rmp}. Although originally proposed to explain the formation of patterns in the macroscopic world, recent studies found that this idea is very much applicable even in the atomic scale and the microscopic quantum domain where the rules are governed by quantum mechanical constraints. For example, \citet{qfluid} reported the appearance of Turing patterns in a coherent quantum fluid of microcavity polaritons. \citet{bismath} provided an experimental evidence of Turing patterns in the atomic scale (e.g., in the atomic bismuth monolayer). The observation of Turing patterns in quantum systems is a promising topic of research as it leads to a broader understanding of Turing patterns and their possible application potentiality \cite{qfluid}.

\citet{kosprl} discovered the Turing bifurcation in ``classical'' coupled oscillators that governs a transition from a homogeneous steady state to an inhomogeneous steady state. The homogeneous steady state is widely known as the amplitude death (AD) state \cite{prasad1}, and the inhomogeneous one is called the oscillation death (OD) state \cite{kosprep}. Unlike the original spatiotemporal Turing scenario, in the case of coupled oscillators the space is discrete and the oscillator index plays the role of the space variable. Later on this significant transition was verified experimentally in electronic circuits and chemical oscillators \cite{tanpre2}. The Turing transition from AD to OD in classical oscillators is generally believed to be relevant in understanding cellular differentiation and other symmetry breaking phenomena in biological systems \cite{kosprep}. A recent burst of publications reported this transition in a variety of classical systems under diverse coupling schemes~\cite{kosprep,kospre,kurthpre,dana,tanpre1,tanpre3,bandutta,acamc}. 

Although many important emergent dynamics of coupled oscillators, such as synchronization \cite{lee_prl,brud_prl1,squeezing,expt1,expt2,schoell_qm} and oscillation suppression \cite{qad1,qad2,qmod,qrev} have been explored in the quantum regime, surprisingly, the quantum analogue of the Turing bifurcation route from  homogeneous to inhomogeneous steady state has remained been unobserved. Recent studies on Pyragas control \cite{schoellqm2}, coherence resonance \cite{qmcores}, and relaxation oscillations \cite{chia,lif} in quantum systems broadened our understanding of the well known dynamical behaviors in the quantum world. A continuous pursuit of unravelling the classical dynamics in the quantum regime motivates us to search for the Turing type bifurcation in coupled quantum oscillators.

The quantum amplitude suppression was reported by \citet{qad1}. They found that unlike the classical AD state, in the quantum AD (QAD) state the quantum noise resists the complete cessation of oscillations. In the quantum domain, QAD is manifested in the pronounced reduction in the bosonic excitations (e.g., phonon or photon). Later, \citet{qad2} found that a Kerr-type nonlinearity enhances the QAD state. However, earlier studies did not observe the quantum inhomogeneous steady state or the quantum OD  (QOD) state. The QOD state has recently been discovered by \citet{qmod} that appears only in the deep quantum regime. In the deep quantum regime the notion of QAD is illusive because in this region the number of bosonic excitations is sparse, and the inherent quantum noise  dominates the dynamics. Although Refs.~\cite{qmod,qrev} reported QOD in the deep quantum regime, the Turing bifurcation route of QAD--QOD transition was not observed there as in this regime the QAD is indistinguishable from a quantum limit cycle. 
Since, the QAD and QOD appear in two different regimes of quantum domain (QAD: weak quantum regime; QOD: deep quantum regime), the direct transition from QAD to QOD through the Turing bifurcation has remained been unobserved. 

In this paper, we discover the quantum analogue of the Turing bifurcation that gives a  transition from QAD to QOD. We found both QAD and QOD and their transition in the weak quantum regime in quantum van der Pol oscillators coupled through a conjugate coupling. The conjugate coupling scheme was proposed by \citet{karnatak} as a general scheme to induce AD. This coupling is relevant in systems where the state of a variable interacts with the outcome of a dissimilar variable; For example, in the case of semiconductor lasers  the photon intensity interacts with the injection current \cite{rajconj}. Here we propose the quantum version of this coupling scheme and apply it on quantum van der Pol oscillators. We constitute the quantum master equation of the coupled system in the Lindblad form. Direct simulation of the master equation gives the evidence of the Turing bifurcation. We further support our results using the analytical treatment of the equivalent noisy classical model.

\section{Coupled van der Pol oscillators}
\label{sec:vdp}
We start with two coupled ``classical'' van der Pol (vdP) oscillators \cite{vdposc} which are coupled via scalar conjugate coupling. A mathematical model of the coupled system reads
\begin{subequations}
\label{vdp}
\begin{equation}
\dot{x_j}=\omega y_j + \ve(y_{j'} - x_j),
\end{equation}
\begin{equation}
\dot{y_j}=-\omega x_j + (k_1-8k_2 {x_j}^2)y_j,
\end{equation}
\end{subequations}
where $j=1, 2$, $j'=1, 2$ and $j\neq j'$. $\omega$ is the intrinsic frequency of each oscillator. Here $k_1$ is the coefficient of linear pumping, $k_2$ is the coefficient of  nonlinear damping ($k_1,k_2>0$), and $\ve$ is the coupling strength. For $\ve=0$, Eq.\eqref{vdp} represents the uncoupled van der Pol oscillators.

\begin{figure}
\includegraphics[width=.4\textwidth]{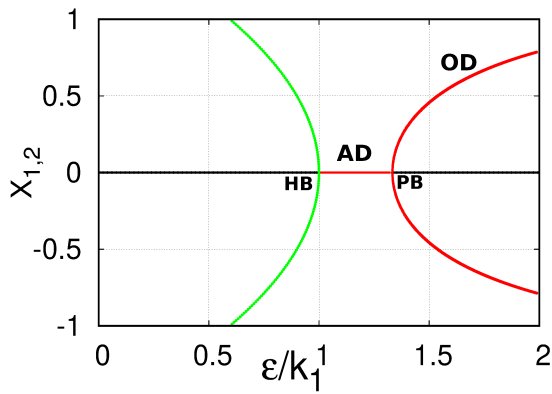}
\caption{Classical bifurcation diagram of \eqref{vdp}. $x_{1,2}$ vs. $\ve/k_1$. It shows a Turing bifurcation route from a homogeneous state (AD) to an inhomogeneous state (OD). HB: Hopf bifurcation, PB: pitchfork bifurcation. Red (grey): stable steady state, black: unstable steady state, green (light grey): stable limit cycle. Other parameters are $\omega=2$ and $(k_1, k_2)=(1, 0.2)$.}
\label{f:clbif}
\end{figure}

Note that \eqref{vdp} has a trivial steady state at the origin, $F_{HSS}\equiv(0,0,0,0)$, and additionally, it has one nontrivial fixed point,  $F_{IHSS}\equiv (x^*,y^*,-x^*,-y^*)$, where $x^*=\frac{\omega-\varepsilon}{\varepsilon}y^*$ and $y^*=\frac{\sqrt{k_1-\frac{\omega(\omega-\varepsilon)}{\varepsilon}}}{\sqrt{8k_2}(1-\frac{\omega}{\varepsilon})}$. A bifurcation analysis shows that, with increasing $\ve$ oscillations cease through an inverse Hopf bifurcation at $\varepsilon_{HB}=k_1$ and AD emerges where all the oscillators arrive at the stable homogeneous steady state (HSS). The HSS or the AD state losses its stability through a supercritical pitchfork bifurcation at $\varepsilon_{PB}=\frac{\omega^2}{\omega+k_1}$ and as a result two inhomogeneous steady states (IHSS) emerge: This is the well known {\it Turing bifurcation} that gives heterogeneity from a homogeneous solution through a spontaneous symmetry-breaking mechanism. Figure~\ref{f:clbif} presents the bifurcation diagram (using XPPAUT \cite{xpp}) with $\ve/k_1$ that clearly shows this scenario (for $\omega=2$, $k_1=1$, and $k_2=0.2$).

We can express Eq.~\eqref{vdp} in terms of a complex amplitude $\alpha_j=x_j+iy_j$ and the corresponding amplitude equation is derived using the harmonic approximation \cite{qmod}, which reads
\begin{equation}
\label{amp}
\begin{split}
\dot{\alpha_j}&=-i\omega\alpha_j+(\frac{k_1}{2}-k_2|\alpha_j|^2)\alpha_j \\
&-\frac{\ve}{2}\left[(\alpha_j+{\alpha_j}^*)+i(\alpha_{j'}-{\alpha_{j'}}^*)\right].
\end{split}
\end{equation}
At $\ve=0$, the uncoupled oscillators show a limit cycle oscillation with an amplitude $\sqrt{\frac{k_1}{2k_2}}$.

The coupled quantum van der Pol oscillators (Eq.\eqref{amp}) can be represented by the quantum master equation in the density matrix $\rho$:
\begin{equation}
\label{master}
\begin{split}
\dot{\rho}&=-i[\omega ({a_1}^\dag a_1+{a_2}^\dag a_2)+\frac{\ve}{2}({a_1}^\dag a_2+{a_2}^\dag a_1)\\
&-\frac{\ve}{2}({a_1}^\dag {a_2}^\dag+a_1a_2)-\frac{i\ve}{4}({{a_1}^\dag}^2+{{a_2}^\dag}^2-{a_1}^2-{a_2}^2),\rho]\\
&+k_1\sum_{j=1}^{2}\mathcal{D}[{a_j}^\dag](\rho)+k_2\sum_{j=1}^{2}\mathcal{D}[{a_j}^2](\rho)+\ve\sum_{j=1}^{2}\mathcal{D}[a_j](\rho),
\end{split}
\end{equation}
where $\mathcal{D}[\hat{L}](\rho)$ is the Lindblad dissipator having the form $\mathcal{D}[\hat{L}](\rho)=\hat{L}\rho \hat{L}^\dag-\frac{1}{2}\{\hat{L}^\dag \hat{L},\rho \}$, where $\hat{L}$ is an operator (without any loss of generality we set $\hbar=1$). $a_j$ and ${a_j}^\dag$ are the bosonic annihilation and creation operators of the $j$-th oscillator, respectively. The implications of $k_1$ and $k_2$ in the quantum regime can be understood from \eqref{master}: $k_1$ is the rate of single photon creation (equivalent to the linear pumping), and $k_2$ gives the rate of two-photon absorption (equivalent to the nonlinear damping). The last term of \eqref{master} gives the coupling-dependent single photon annihilation whose rate is determined by the coupling strength $\ve$ ($\ge 0$). Intuitively, this last term induces an additional loss of photons that may result in oscillation suppression. 

In the semiclassical limit, the linear pumping rate dominates over the nonlinear damping rate, i.e., $k_1>k_2$ and one may approximate $\langle a\rangle \equiv \alpha$. Under this condition, the quantum master equation \eqref{master} is equivalent to the classical amplitude equation \eqref{amp} by the following relation \cite{qmod}: $\dot{\braket{a}}=\mbox{Tr}(\dot \rho a)$.

\section{Results}\label{sec:results}
\subsection{Simulation of quantum master equation}
We numerically solve the master equation \eqref{master} using a Python based quantum simulator package \texttt{QuTiP} \cite{qutip}. To visualize the dynamics of the coupled system we compute the Wigner function using the steady state solution of the density matrix \cite{knightbook}. The Wigner function is a quasi-probability distribution function that provides a reliable picture of the quantum dynamics \cite{squeezing}. Moreover, it is also accessible in experimental set ups \cite{wigner}. In the numerical simulations we choose the following parameters: $\omega=2$ and $(k_1, k_2)=(1, 0.2)$. Note that $k_1>k_2$ ensures that the system resides in the weak quantum regime where semiclassical treatments are applicable and the QAD is distinguishable from a quantum limit cycle. 

\begin{figure}
\includegraphics[width=.48\textwidth]{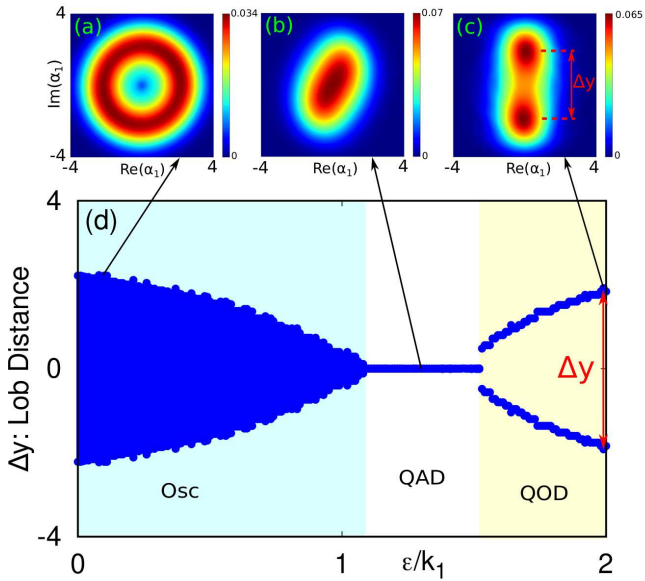}
\caption{The Wigner function at (a) $\ve=0.1$: quantum oscillatory state, (b) $\ve=1.3$: quantum amplitude death (QAD) state, and (c) $\ve=1.99$: quantum oscillation death (QOD) state. (d) Quantum Turing bifurcation: The variation of the $y$-component of local maximum values of the Wigner function ($\Delta y$) plotted with $\ve/k_1$. Note the transition from QAD to QOD through the quantum Turing bifurcation. Other parameters are $\omega=2$ and $(k_1, k_2)=(1, 0.2)$.}
\label{quantum}
\end{figure}

\begin{figure}[t]
\includegraphics[width=.48\textwidth]{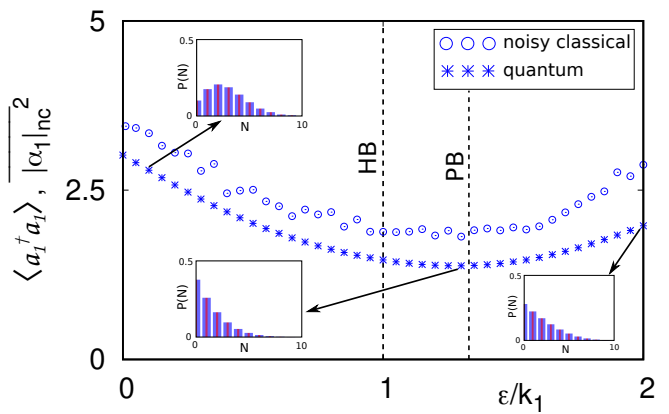}
\caption{Mean photon number from the quantum model ($\braket{a_1^\dag a_1}$) and the averaged amplitude from the noisy classical model ($\overline{{|\alpha _1|_{nc}}^2}$). Insets show the occupation of the Fock levels for quantum limit cycle $\ve=0.1$, QAD ($\ve=1.3$) and QOD ($\ve=1.99$). Other parameters are $\omega=2$ and $(k_1, k_2)=(1, 0.2)$.}
\label{semiclassical}
\end{figure}

We plot the Wigner function at three exemplary values of $\ve$. At $\ve=0.1$, Fig.~\ref{quantum}(a) demonstrates a quantum limit cycle that is visible from the ring shaped Wigner function. Fig.\ref{quantum}(b) presents the same but at a higher coupling constant $\ve=1.3$: it depicts a QAD state as now the Wigner function shows the maximum probability at the origin. Note that unlike the classical AD, here the complete cessation of oscillation is prohibited due to the presence of inherent quantum noise. Also, significantly, the QAD state is a squeezed quantum state {\it that has no counterpart in the classical dynamics}. At a stronger coupling strength, the QAD state is transformed into a quantum OD (QOD) state, which is depicted by a bimodal Wigner function: see Fig.\ref{quantum}(c) for $\ve=1.99$. Here the maximum probability is shifted from the origin and two prominent lobs are appeared representing two branches of the inhomogeneous steady state. Note that, in spite of the quantum noise which tends to homogenize the inhomogeneous steady states, the lobs are distinguishable.   

To show that this transition from QAD to QOD is continuous we define a variable $\Delta y$, which is the distance between the local maxima of the projection of the Wigner function on the $y$-axis (see Fig.~\ref{quantum}(c)). Figure~\ref{quantum}(d) demonstrates the variation of $\Delta y$ with increasing $\ve$. The oscillatory zone (indicated by Osc) shows a decreasing amplitude with an increasing coupling strength. Beyond a certain $\ve$, the QAD appears, where we have only a single local maxima (i.e., $\Delta y=0$). Further increase of coupling strength results in a transition from QAD to QOD where the unimodal Wigner function splits into a bimodal shape whose lobs are separated by $\Delta y \ne 0$. Remarkably, the parametric zone of the QAD and QOD qualitatively matches with the corresponding classical counterparts (cf. Fig.~\ref{f:clbif}).

The occurrence of oscillation suppression can be better understood from the mean photon number, $\braket{{a_1}^\dag a_1}$. This is shown in Fig.~\ref{semiclassical}: it shows a decreasing mean photon number with increasing $\ve/k_1$ indicating the suppression of oscillation. Due to the quantum noise, $\braket{{a_1}^\dag a_1}$ does not reach to zero, however, it shows a moderate collapse of the oscillation. The mean photon number attains lower values between $\ve_{HB}$ and $\ve_{PB}$ (as calculated in Sec.~\ref{sec:vdp} for the classical case) indicating the onset of QAD. The occupation of the corresponding Fock states for exemplary $\ve/k_1$ values are also shown in the inset. It can be seen that in the quantum oscillatory state ($\ve/k_1$=0.1) the higher Fock levels are populated, however, in the QAD ($\ve/k_1$=1.3) and QOD states ($\ve/k_1=1.99$) only the lowest lying Fock levels are populated. 

Therefore, the observed scenario of the transition from QAD to QOD states has a striking one to one resemblance with the classical Turing bifurcation and clearly provides evidence to  the quantum version of the Turing bifurcation.


\subsection{Noisy classical model: Analysis}\label{sec:noisyclassical} 

To further support the observed quantum Turing bifurcation [Fig.\ref{quantum}(a-d)], we compare the quantum scenarios with the corresponding noisy classical model (or the semiclassical model). It will give a conclusive  analytical evidence of the observed behavior in the presence of noise whose intensity is equal to the inherent quantum noise \cite{qad1}. We evaluate the intensity of quantum noise from the quantum master equation following Ref.~\cite{qad1}. 
The quantum master equation \eqref{master} is represented in the phase space using a  partial differential equation of the Wigner distribution function ($W(\bm{\alpha})$) \cite{carmichael} that reads
\begin{equation}
\label{diff_eqn_w}
\begin{split}
\partial_t{W(\bm{\alpha})}&=\sum_{j=1}^2\left[-\left(\frac{\partial}{\partial \alpha _j}\mu _{\alpha _j}+c.c.\right) \right. \\
&+ \left. \frac{1}{2}\left(\frac{\partial ^2}{\partial \alpha _j \partial {\alpha_j}^*}D_{\alpha _j {\alpha_j}^*}+\frac{\partial ^2}{\partial \alpha _j \partial {\alpha_{j'}}^*}D_{\alpha _j {\alpha_{j'}}^*} \right) \right. \\
&+ \left. \frac{k_2}{4}\left(\frac{\partial ^3}{\partial {\alpha_j}^* \partial {\alpha_j}^2}\alpha _j+c.c\right) \right]W(\bm{\alpha}),
\end{split}
\end{equation}
where the elements of the drift vector ($\bm{\mu}$) are: $\mu _{\alpha _j}=\left[-i\omega+\frac{k_1}{2}-k_2(|\alpha _j|^2-1)-\frac{\ve}{2}\right]\alpha _j-\frac{\ve}{2}{\alpha_{j}}^*-\frac{i\ve}{2}{\alpha_{j'}}+\frac{i\ve}{2}{\alpha_{j'}}^*,$
and the elements of the diffusion matrix $\bm{D}$ are:
$D_{\alpha _j {\alpha_j}^*}=k_1+2k_2(2|\alpha _j|^2-1)+\ve, D_{\alpha _j {\alpha_{j'}}^*}=0$, with $j=1,2$, $j'=1,2$ and $j\neq j'$. In the weak nonlinear regime ($k_2\ll k_1$), Eq.~\eqref{diff_eqn_w} reduces to the Fokker-Planck equation, which is given by 
\begin{equation}
\label{fp}
\begin{split}
\partial_t{W}(\textbf{X})&=\sum_{j=1}^2\left[-\left(\frac{\partial}{\partial x_j}\mu _{x_j}+\frac{\partial}{\partial y_j}\mu _{y_j}\right) \right) \\
&+ \left. \frac{1}{2}\left(\frac{\partial ^2}{\partial x_j \partial x_j}D_{x_j x_j}+\frac{\partial ^2}{\partial y_j \partial y_j}D_{y_j y_j} \right. \right. \\
&+ \left. \left. \frac{\partial ^2}{\partial x_j \partial x_{j'}}D_{x_j x_{j'}}+\frac{\partial ^2}{\partial y_j \partial y_{j'}}D_{y_j y_{j'}}  \right) \right]W(\textbf{X}),
\end{split}
\end{equation}
where $\textbf{X}=(x_1, y_1, x_2, y_2)$. The elements of drift vector are,
\begin{subequations}
\label{drift}
\begin{align}
\begin{split}
\mu _{x_j}&=\omega y_j + \left[\frac{k_1}{2}-k_2({x_j}^2+{y_j}^2-1)-\ve\right]x_j \\
&+ \ve y_{j'},
\end{split} \\
\begin{split}
\mu _{y_j}&=-\omega x_j + \left[\frac{k_1}{2}-k_2({x_j}^2+{y_j}^2-1)\right]y_j.
\end{split}
\end{align}
\end{subequations}
The diffusion matrix has the following form,
\begin{equation}\label{diffusion_mat} 
{\bm{D}}=\frac{1}{2}\left(\begin{array}{cccc} \nu _1 & 0 & 0 & 0 \\
0 & \nu _1 & 0 & 0\\
0 & 0 & \nu _2 & 0 \\
 0 & 0 & 0 & \nu _2  \end{array}\right).
\end{equation}
where $\nu _j=\frac{k_1}{2}+k_2[2({x_j}^2+{y_j}^2)-1]+\frac{\ve}{2}$.
From Eq.\eqref{fp}, the following stochastic differential equation can be derived,
\begin{equation}\label{sde}
d\textbf{X}=\bm{\mu}dt+\bm{\sigma} d\textbf{W}_t,
\end{equation}
where $\bm{\sigma}$ is the noise strength and $d\textbf{W}_t$ is the Wiener increment. As the diffusion matrix $\bm{D}$ (given in Eq.\eqref{diffusion_mat}) is diagonal, we can analytically derive $\bm{\sigma}$ from it as $\bm{\sigma}=\sqrt{\bm{D}}$.

By solving the stochastic differential equation (Eq.~\eqref{sde}) (using JiTCSDE module in Python \cite{jitcode}), we compute the behaviour of the noisy-classical system starting from random initial conditions (we take 1000 independent run). The average amplitude from the noisy classical model ($\overline{{|\alpha _1|_{nc}}^2}$) with $\ve/k_1$ is demonstrated in Fig.~\ref{semiclassical}. It shows a decreasing amplitude with increasing coupling strength resembling the quantum scenario. Note that the mean photon number, $\braket{{a_1}^\dag a_1}$, of the quantum case always lies below $\overline{{|\alpha _1|_{nc}}^2}$, which is a general indicator of the quantum oscillation suppression scenario \cite{qad1,qmod,qrev}.  

\begin{figure}[t]
\includegraphics[width=.48\textwidth]{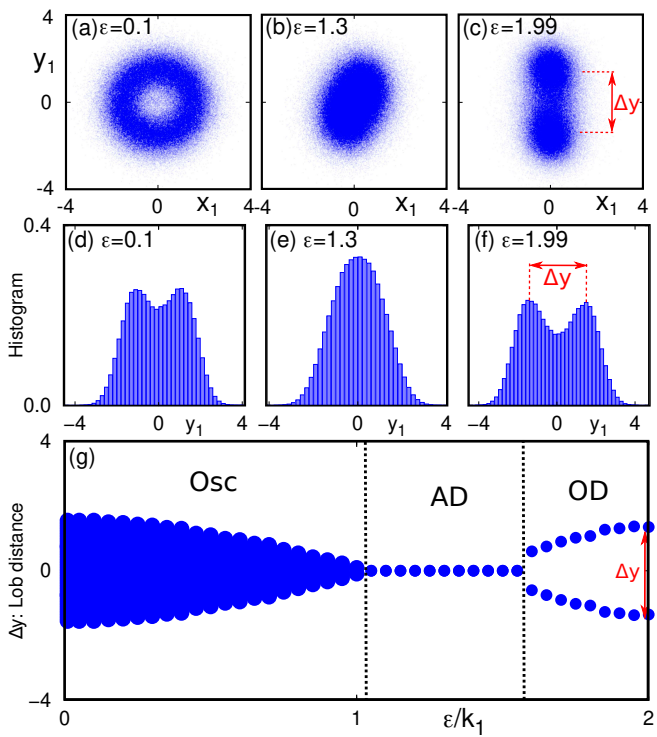}
\caption{{\bf Noisy classical case}: Phase space plot for (a) $\ve=0.1$: noisy limit cycle, (b) $\ve=1.3$: noisy AD and (c) $\ve=1.99$: noisy QOD. The corresponding histograms are shown in (d)--(f). (g) Bifurcation diagram using the local maxima of the phase trajectories showing a Turing bifurcation route from noisy AD to OD. Note that in the OD state, $\Delta y$ gives the distance between two local maxima of the histogram. Other parameters are $\omega=2$ and $(k_1, k_2)=(1, 0.2)$.}
\label{f:semicl}
\end{figure}

Figures~\ref{f:semicl}(a--g) summarize the results of the noisy classical model. The noisy limit cycle is demonstrated in the phase space in Fig.~\ref{f:semicl}(a) at $\ve=0.1$. A weighted histogram of the same is shown in Fig.~\ref{f:semicl}(d). The scenario at a larger coupling strength, $\ve=1.3$, is shown in Fig.~\ref{f:semicl}(b, e). Fig.~\ref{f:semicl}(b) demonstrates that in the phase space the phase points are crowded around the origin and Fig.~\ref{f:semicl}(e) presents the corresponding histogram, which obeys a Gaussian distribution. Both of these indicate the occurrence of a noisy AD. Finally, Fig.~\ref{f:semicl}(c) shows the appearance of an inhomogeneous steady state or the noisy OD at a higher coupling strength, $\ve=1.99$: two distinct but noisy lobs correspond to two different branches of the OD. The corresponding histogram [Fig.~\ref{f:semicl}(f)] exhibits a two hump nature indicating the occurrence of two lobs.

We draw a representative bifurcation diagram in Fig.~\ref{f:semicl}(g) showing the local maxima of the averaged phase trajectories. In the case of a noisy AD state there is only one local maxima situated at zero. However, in the case of a noisy OD state two local maxima are separated by a distance $\Delta y$ ($ne 0$) in the $y$-axis. They are equivalent to the two lobs/branches of the QOD. 
Note the striking resemblance of noisy classical results [Fig.~\ref{f:semicl}] with the quantum results [Fig.~\ref{quantum}]. Both the quantum and noisy classical results confirm the occurrence of the Turing bifurcation route from homogeneous to inhomogeneous steady states in coupled quantum oscillators.

\section{Conclusions}
\label{sec:con}
In this paper we have reported a transition from a homogeneous steady state (QAD) to an  inhomogeneous steady state (QOD) via the quantum Turing bifurcation in coupled quantum oscillators. The direct simulation of the quantum master equation and analytical investigations through the noisy classical model of the coupled system established the appearance of the quantum Turing bifurcation. Unlike earlier studies \cite{qmod,qrev}, where QOD was found to appear in the deep quantum regime only, here we have observed both QAD and QOD, and their transition in the weak quantum regime. This enable us to give conclusive evidence of the Turing bifurcation in the quantum regime as now the notion of both QAD and QOD are unambiguous. Our study reveals that the Turing bifurcation scenario of transition from homogeneous to inhomogeneous steady states as reported in Ref.~\cite{kosprl} for classical oscillators is indeed a general phenomenon and occurs even in the quantum regime.

We believe that the recent advancement of experimental techniques would enable one to implement our coupled system in experimental set ups, such as, the trapped-ion \cite{lee_prl,expt-ion} and the opto-mechanical experiments \cite{expt-mem}. Further, the symmetry breaking mechanism studied here can be extended to networks of quantum oscillators to explore the possibility of partial synchronization and chimera death states \cite{scholl-CD,annabook,sl-front,tanCD,acamc} in the quantum regime. The present study significantly broadens our understanding of the symmetry-breaking mechanism in the interface of quantum-classical domain. Moreover, the quantum Turing mechanism can be explored in the light of harnessing its potential in applications like quantum computations and cryptography \cite{qfluid}.

\begin{acknowledgments}
B.B. and T.K acknowledge the financial assistance from the University Grants Commission (UGC), India. T. B. acknowledges the financial support from the Science and Engineering Research Board (SERB), Government of India, in the form of a Core Research Grant [CRG/2019/002632].
\end{acknowledgments}


\begin{thebibliography}{45}%
\makeatletter
\providecommand \@ifxundefined [1]{%
 \@ifx{#1\undefined}
}%
\providecommand \@ifnum [1]{%
 \ifnum #1\expandafter \@firstoftwo
 \else \expandafter \@secondoftwo
 \fi
}%
\providecommand \@ifx [1]{%
 \ifx #1\expandafter \@firstoftwo
 \else \expandafter \@secondoftwo
 \fi
}%
\providecommand \natexlab [1]{#1}%
\providecommand \enquote  [1]{``#1''}%
\providecommand \bibnamefont  [1]{#1}%
\providecommand \bibfnamefont [1]{#1}%
\providecommand \citenamefont [1]{#1}%
\providecommand \href@noop [0]{\@secondoftwo}%
\providecommand \href [0]{\begingroup \@sanitize@url \@href}%
\providecommand \@href[1]{\@@startlink{#1}\@@href}%
\providecommand \@@href[1]{\endgroup#1\@@endlink}%
\providecommand \@sanitize@url [0]{\catcode `\\12\catcode `\$12\catcode
  `\&12\catcode `\#12\catcode `\^12\catcode `\_12\catcode `\%12\relax}%
\providecommand \@@startlink[1]{}%
\providecommand \@@endlink[0]{}%
\providecommand \url  [0]{\begingroup\@sanitize@url \@url }%
\providecommand \@url [1]{\endgroup\@href {#1}{\urlprefix }}%
\providecommand \urlprefix  [0]{URL }%
\providecommand \Eprint [0]{\href }%
\providecommand \doibase [0]{https://doi.org/}%
\providecommand \selectlanguage [0]{\@gobble}%
\providecommand \bibinfo  [0]{\@secondoftwo}%
\providecommand \bibfield  [0]{\@secondoftwo}%
\providecommand \translation [1]{[#1]}%
\providecommand \BibitemOpen [0]{}%
\providecommand \bibitemStop [0]{}%
\providecommand \bibitemNoStop [0]{.\EOS\space}%
\providecommand \EOS [0]{\spacefactor3000\relax}%
\providecommand \BibitemShut  [1]{\csname bibitem#1\endcsname}%
\let\auto@bib@innerbib\@empty
\bibitem [{\citenamefont {Turing}(1952)}]{turing}%
  \BibitemOpen
  \bibfield  {author} {\bibinfo {author} {\bibfnamefont {A.}~\bibnamefont
  {Turing}},\ }\bibfield  {title} {\bibinfo {title} {The chemical basis of
  morphogenesis},\ }\href@noop {} {\bibfield  {journal} {\bibinfo  {journal}
  {Philos. Trans. R. Soc. Lond.}\ }\textbf {\bibinfo {volume} {237}},\ \bibinfo
  {pages} {37} (\bibinfo {year} {1952})}\BibitemShut {NoStop}%
\bibitem [{\citenamefont {Murray}(2003)}]{murry2}%
  \BibitemOpen
  \bibfield  {author} {\bibinfo {author} {\bibfnamefont {J.~D.}\ \bibnamefont
  {Murray}},\ }\href@noop {} {\emph {\bibinfo {title} {Mathematical Biology
  {II}}}}\ (\bibinfo  {publisher} {Springer-Verlag},\ \bibinfo {address} {New
  York},\ \bibinfo {year} {2003})\BibitemShut {NoStop}%
\bibitem [{\citenamefont {Koch}\ and\ \citenamefont
  {Meinhardt}(1994)}]{tur-rmp}%
  \BibitemOpen
  \bibfield  {author} {\bibinfo {author} {\bibfnamefont {A.~J.}\ \bibnamefont
  {Koch}}\ and\ \bibinfo {author} {\bibfnamefont {H.}~\bibnamefont
  {Meinhardt}},\ }\bibfield  {title} {\bibinfo {title} {Biological pattern
  formation: from basic mechanisms to complex structures},\ }\href@noop {}
  {\bibfield  {journal} {\bibinfo  {journal} {Rev. Mod. Phys.}\ }\textbf
  {\bibinfo {volume} {66}},\ \bibinfo {pages} {1481} (\bibinfo {year}
  {1994})}\BibitemShut {NoStop}%
\bibitem [{\citenamefont {Ardizzone}\ \emph {et~al.}(2013)\citenamefont
  {Ardizzone}, \citenamefont {Lewandowski}, \citenamefont {Luk}, \citenamefont
  {Tse}, \citenamefont {Kwong}, \citenamefont {L{\"u}cke}, \citenamefont
  {Abbarchi}, \citenamefont {Baudin}, \citenamefont {Galopin}, \citenamefont
  {Bloch}, \citenamefont {Lemaitre}, \citenamefont {Leung}, \citenamefont
  {Roussignol}, \citenamefont {Binder}, \citenamefont {Tignon},\ and\
  \citenamefont {Schumacher}}]{qfluid}%
  \BibitemOpen
  \bibfield  {author} {\bibinfo {author} {\bibfnamefont {V.}~\bibnamefont
  {Ardizzone}}, \bibinfo {author} {\bibfnamefont {P.}~\bibnamefont
  {Lewandowski}}, \bibinfo {author} {\bibfnamefont {M.~H.}\ \bibnamefont
  {Luk}}, \bibinfo {author} {\bibfnamefont {Y.~C.}\ \bibnamefont {Tse}},
  \bibinfo {author} {\bibfnamefont {N.~H.}\ \bibnamefont {Kwong}}, \bibinfo
  {author} {\bibfnamefont {A.}~\bibnamefont {L{\"u}cke}}, \bibinfo {author}
  {\bibfnamefont {M.}~\bibnamefont {Abbarchi}}, \bibinfo {author}
  {\bibfnamefont {E.}~\bibnamefont {Baudin}}, \bibinfo {author} {\bibfnamefont
  {E.}~\bibnamefont {Galopin}}, \bibinfo {author} {\bibfnamefont
  {J.}~\bibnamefont {Bloch}}, \bibinfo {author} {\bibfnamefont
  {A.}~\bibnamefont {Lemaitre}}, \bibinfo {author} {\bibfnamefont {P.~T.}\
  \bibnamefont {Leung}}, \bibinfo {author} {\bibfnamefont {P.}~\bibnamefont
  {Roussignol}}, \bibinfo {author} {\bibfnamefont {R.}~\bibnamefont {Binder}},
  \bibinfo {author} {\bibfnamefont {J.}~\bibnamefont {Tignon}},\ and\ \bibinfo
  {author} {\bibfnamefont {S.}~\bibnamefont {Schumacher}},\ }\bibfield  {title}
  {\bibinfo {title} {Formation and control of turing patterns in a coherent
  quantum fluid},\ }\href@noop {} {\bibfield  {journal} {\bibinfo  {journal}
  {Sci. Rep.}\ }\textbf {\bibinfo {volume} {3}},\ \bibinfo {pages} {3016}
  (\bibinfo {year} {2013})}\BibitemShut {NoStop}%
\bibitem [{\citenamefont {Fuseya}\ \emph {et~al.}(2021)\citenamefont {Fuseya},
  \citenamefont {Katsuno}, \citenamefont {Behnia},\ and\ \citenamefont
  {Kapitulnik}}]{bismath}%
  \BibitemOpen
  \bibfield  {author} {\bibinfo {author} {\bibfnamefont {Y.}~\bibnamefont
  {Fuseya}}, \bibinfo {author} {\bibfnamefont {H.}~\bibnamefont {Katsuno}},
  \bibinfo {author} {\bibfnamefont {K.}~\bibnamefont {Behnia}},\ and\ \bibinfo
  {author} {\bibfnamefont {A.}~\bibnamefont {Kapitulnik}},\ }\href@noop {}
  {\bibinfo {title} {Nanometric turing patterns: Morphogenesis of a bismuth
  monolayer}} (\bibinfo {year} {2021}),\ \Eprint
  {https://arxiv.org/abs/2104.01058} {arXiv:2104.01058 [cond-mat.mes-hall]}
  \BibitemShut {NoStop}%
\bibitem [{\citenamefont {Koseska}\ \emph
  {et~al.}(2013{\natexlab{a}})\citenamefont {Koseska}, \citenamefont {Volkov},\
  and\ \citenamefont {Kurths}}]{kosprl}%
  \BibitemOpen
  \bibfield  {author} {\bibinfo {author} {\bibfnamefont {A.}~\bibnamefont
  {Koseska}}, \bibinfo {author} {\bibfnamefont {E.}~\bibnamefont {Volkov}},\
  and\ \bibinfo {author} {\bibfnamefont {J.}~\bibnamefont {Kurths}},\
  }\bibfield  {title} {\bibinfo {title} {Transition from amplitude to
  oscillation death via turing bifurcation},\ }\href@noop {} {\bibfield
  {journal} {\bibinfo  {journal} {Phys. Rev. Lett}\ }\textbf {\bibinfo {volume}
  {111}},\ \bibinfo {pages} {024103} (\bibinfo {year}
  {2013}{\natexlab{a}})}\BibitemShut {NoStop}%
\bibitem [{\citenamefont {Saxena}\ \emph {et~al.}(2012)\citenamefont {Saxena},
  \citenamefont {Prasad},\ and\ \citenamefont {Ramaswamy}}]{prasad1}%
  \BibitemOpen
  \bibfield  {author} {\bibinfo {author} {\bibfnamefont {G.}~\bibnamefont
  {Saxena}}, \bibinfo {author} {\bibfnamefont {A.}~\bibnamefont {Prasad}},\
  and\ \bibinfo {author} {\bibfnamefont {R.}~\bibnamefont {Ramaswamy}},\
  }\bibfield  {title} {\bibinfo {title} {Amplitude death: The emergence of
  stationarity in coupled nonlinear systems},\ }\href@noop {} {\bibfield
  {journal} {\bibinfo  {journal} {Physics Reports}\ }\textbf {\bibinfo {volume}
  {521}},\ \bibinfo {pages} {205} (\bibinfo {year} {2012})}\BibitemShut
  {NoStop}%
\bibitem [{\citenamefont {Koseska}\ \emph
  {et~al.}(2013{\natexlab{b}})\citenamefont {Koseska}, \citenamefont {Volkov},\
  and\ \citenamefont {Kurths}}]{kosprep}%
  \BibitemOpen
  \bibfield  {author} {\bibinfo {author} {\bibfnamefont {A.}~\bibnamefont
  {Koseska}}, \bibinfo {author} {\bibfnamefont {E.}~\bibnamefont {Volkov}},\
  and\ \bibinfo {author} {\bibfnamefont {J.}~\bibnamefont {Kurths}},\
  }\bibfield  {title} {\bibinfo {title} {Oscillation quenching mechanisms:
  Amplitude vs oscillation death},\ }\href@noop {} {\bibfield  {journal}
  {\bibinfo  {journal} {Phys. Reports}\ }\textbf {\bibinfo {volume} {531}},\
  \bibinfo {pages} {173} (\bibinfo {year} {2013}{\natexlab{b}})}\BibitemShut
  {NoStop}%
\bibitem [{\citenamefont {Banerjee}\ and\ \citenamefont
  {Ghosh}(2014{\natexlab{a}})}]{tanpre2}%
  \BibitemOpen
  \bibfield  {author} {\bibinfo {author} {\bibfnamefont {T.}~\bibnamefont
  {Banerjee}}\ and\ \bibinfo {author} {\bibfnamefont {D.}~\bibnamefont
  {Ghosh}},\ }\bibfield  {title} {\bibinfo {title} {Experimental observation of
  a transition from amplitude to oscillation death in coupled oscillators},\
  }\href@noop {} {\bibfield  {journal} {\bibinfo  {journal} {Phys. Rev. E}\
  }\textbf {\bibinfo {volume} {89}},\ \bibinfo {pages} {062902} (\bibinfo
  {year} {2014}{\natexlab{a}})}\BibitemShut {NoStop}%
\bibitem [{\citenamefont {Zou}\ \emph {et~al.}(2013)\citenamefont {Zou},
  \citenamefont {Senthilkumar}, \citenamefont {Koseska},\ and\ \citenamefont
  {Kurths}}]{kospre}%
  \BibitemOpen
  \bibfield  {author} {\bibinfo {author} {\bibfnamefont {W.}~\bibnamefont
  {Zou}}, \bibinfo {author} {\bibfnamefont {D.~V.}\ \bibnamefont
  {Senthilkumar}}, \bibinfo {author} {\bibfnamefont {A.}~\bibnamefont
  {Koseska}},\ and\ \bibinfo {author} {\bibfnamefont {J.}~\bibnamefont
  {Kurths}},\ }\bibfield  {title} {\bibinfo {title} {Generalizing the
  transition from amplitude to oscillation death in coupled oscillators},\
  }\href@noop {} {\bibfield  {journal} {\bibinfo  {journal} {Phys. Rev. E}\
  }\textbf {\bibinfo {volume} {88}},\ \bibinfo {pages} {050901(R)} (\bibinfo
  {year} {2013})}\BibitemShut {NoStop}%
\bibitem [{\citenamefont {Zou}\ \emph {et~al.}(2014)\citenamefont {Zou},
  \citenamefont {Senthilkumar}, \citenamefont {Duan},\ and\ \citenamefont
  {Kurths}}]{kurthpre}%
  \BibitemOpen
  \bibfield  {author} {\bibinfo {author} {\bibfnamefont {W.}~\bibnamefont
  {Zou}}, \bibinfo {author} {\bibfnamefont {D.~V.}\ \bibnamefont
  {Senthilkumar}}, \bibinfo {author} {\bibfnamefont {J.}~\bibnamefont {Duan}},\
  and\ \bibinfo {author} {\bibfnamefont {J.}~\bibnamefont {Kurths}},\
  }\bibfield  {title} {\bibinfo {title} {Emergence of amplitude and oscillation
  death in identical coupled oscillators},\ }\href@noop {} {\bibfield
  {journal} {\bibinfo  {journal} {Phys. Rev. E}\ }\textbf {\bibinfo {volume}
  {90}},\ \bibinfo {pages} {032906} (\bibinfo {year} {2014})}\BibitemShut
  {NoStop}%
\bibitem [{\citenamefont {Hens}\ \emph {et~al.}(2013)\citenamefont {Hens},
  \citenamefont {Olusola}, \citenamefont {Pal},\ and\ \citenamefont
  {Dana}}]{dana}%
  \BibitemOpen
  \bibfield  {author} {\bibinfo {author} {\bibfnamefont {C.~R.}\ \bibnamefont
  {Hens}}, \bibinfo {author} {\bibfnamefont {O.~I.}\ \bibnamefont {Olusola}},
  \bibinfo {author} {\bibfnamefont {P.}~\bibnamefont {Pal}},\ and\ \bibinfo
  {author} {\bibfnamefont {S.~K.}\ \bibnamefont {Dana}},\ }\bibfield  {title}
  {\bibinfo {title} {Oscillation death in diffusively coupled oscillators by
  local repulsive link},\ }\href@noop {} {\bibfield  {journal} {\bibinfo
  {journal} {Phys. Rev. E}\ }\textbf {\bibinfo {volume} {88}},\ \bibinfo
  {pages} {034902} (\bibinfo {year} {2013})}\BibitemShut {NoStop}%
\bibitem [{\citenamefont {Banerjee}\ and\ \citenamefont
  {Ghosh}(2014{\natexlab{b}})}]{tanpre1}%
  \BibitemOpen
  \bibfield  {author} {\bibinfo {author} {\bibfnamefont {T.}~\bibnamefont
  {Banerjee}}\ and\ \bibinfo {author} {\bibfnamefont {D.}~\bibnamefont
  {Ghosh}},\ }\bibfield  {title} {\bibinfo {title} {Transition from amplitude
  to oscillation death under mean-field diffusive coupling},\ }\href@noop {}
  {\bibfield  {journal} {\bibinfo  {journal} {Phys. Rev. E}\ }\textbf {\bibinfo
  {volume} {89}},\ \bibinfo {pages} {052912} (\bibinfo {year}
  {2014}{\natexlab{b}})}\BibitemShut {NoStop}%
\bibitem [{\citenamefont {Ghosh}\ and\ \citenamefont
  {Banerjee}(2014)}]{tanpre3}%
  \BibitemOpen
  \bibfield  {author} {\bibinfo {author} {\bibfnamefont {D.}~\bibnamefont
  {Ghosh}}\ and\ \bibinfo {author} {\bibfnamefont {T.}~\bibnamefont
  {Banerjee}},\ }\bibfield  {title} {\bibinfo {title} {Transitions among the
  diverse oscillation quenching states induced by the interplay of direct and
  indirect coupling},\ }\href@noop {} {\bibfield  {journal} {\bibinfo
  {journal} {Phys. Rev. E}\ }\textbf {\bibinfo {volume} {90}},\ \bibinfo
  {pages} {062908} (\bibinfo {year} {2014})}\BibitemShut {NoStop}%
\bibitem [{\citenamefont {Banerjee}\ \emph {et~al.}(2015)\citenamefont
  {Banerjee}, \citenamefont {Dutta},\ and\ \citenamefont {Gupta}}]{bandutta}%
  \BibitemOpen
  \bibfield  {author} {\bibinfo {author} {\bibfnamefont {T.}~\bibnamefont
  {Banerjee}}, \bibinfo {author} {\bibfnamefont {P.~S.}\ \bibnamefont
  {Dutta}},\ and\ \bibinfo {author} {\bibfnamefont {A.}~\bibnamefont {Gupta}},\
  }\bibfield  {title} {\bibinfo {title} {Mean-field dispersion-induced spatial
  synchrony, oscillation and amplitude death, and temporal stability in an
  ecological model},\ }\href@noop {} {\bibfield  {journal} {\bibinfo  {journal}
  {Phys. Rev. E}\ }\textbf {\bibinfo {volume} {91}},\ \bibinfo {pages} {052919}
  (\bibinfo {year} {2015})}\BibitemShut {NoStop}%
\bibitem [{\citenamefont {Banerjee}\ \emph {et~al.}(2018)\citenamefont
  {Banerjee}, \citenamefont {Biswas}, \citenamefont {Ghosh}, \citenamefont
  {Sch{\"o}ll},\ and\ \citenamefont {Zakharova}}]{acamc}%
  \BibitemOpen
  \bibfield  {author} {\bibinfo {author} {\bibfnamefont {T.}~\bibnamefont
  {Banerjee}}, \bibinfo {author} {\bibfnamefont {D.}~\bibnamefont {Biswas}},
  \bibinfo {author} {\bibfnamefont {D.}~\bibnamefont {Ghosh}}, \bibinfo
  {author} {\bibfnamefont {E.}~\bibnamefont {Sch{\"o}ll}},\ and\ \bibinfo
  {author} {\bibfnamefont {A.}~\bibnamefont {Zakharova}},\ }\bibfield  {title}
  {\bibinfo {title} {Networks of coupled oscillators: from phase to amplitude
  chimeras},\ }\href@noop {} {\bibfield  {journal} {\bibinfo  {journal}
  {Chaos}\ }\textbf {\bibinfo {volume} {28}},\ \bibinfo {pages} {113124}
  (\bibinfo {year} {2018})}\BibitemShut {NoStop}%
\bibitem [{\citenamefont {Lee}\ and\ \citenamefont
  {Sadeghpour}(2013)}]{lee_prl}%
  \BibitemOpen
  \bibfield  {author} {\bibinfo {author} {\bibfnamefont {T.~E.}\ \bibnamefont
  {Lee}}\ and\ \bibinfo {author} {\bibfnamefont {H.~R.}\ \bibnamefont
  {Sadeghpour}},\ }\bibfield  {title} {\bibinfo {title} {Quantum
  synchronization of quantum van der pol oscillators with trapped ions},\
  }\href@noop {} {\bibfield  {journal} {\bibinfo  {journal} {Phys. Rev. Lett.}\
  }\textbf {\bibinfo {volume} {111}},\ \bibinfo {pages} {234101} (\bibinfo
  {year} {2013})}\BibitemShut {NoStop}%
\bibitem [{\citenamefont {Walter}\ \emph {et~al.}(2014)\citenamefont {Walter},
  \citenamefont {Nunnenkamp},\ and\ \citenamefont {Bruder}}]{brud_prl1}%
  \BibitemOpen
  \bibfield  {author} {\bibinfo {author} {\bibfnamefont {S.}~\bibnamefont
  {Walter}}, \bibinfo {author} {\bibfnamefont {A.}~\bibnamefont {Nunnenkamp}},\
  and\ \bibinfo {author} {\bibfnamefont {C.}~\bibnamefont {Bruder}},\
  }\bibfield  {title} {\bibinfo {title} {Quantum synchronization of a driven
  self-sustained oscillator},\ }\href@noop {} {\bibfield  {journal} {\bibinfo
  {journal} {Phys. Rev. Lett.}\ }\textbf {\bibinfo {volume} {112}},\ \bibinfo
  {pages} {094102} (\bibinfo {year} {2014})}\BibitemShut {NoStop}%
\bibitem [{\citenamefont {Sonar}\ \emph {et~al.}(2018)\citenamefont {Sonar},
  \citenamefont {Hajdu{\v{s}}ek}, \citenamefont {Mukherjee}, \citenamefont
  {Fazio}, \citenamefont {Vedral}, \citenamefont {Vinjanampathy},\ and\
  \citenamefont {Kwek}}]{squeezing}%
  \BibitemOpen
  \bibfield  {author} {\bibinfo {author} {\bibfnamefont {S.}~\bibnamefont
  {Sonar}}, \bibinfo {author} {\bibfnamefont {M.}~\bibnamefont
  {Hajdu{\v{s}}ek}}, \bibinfo {author} {\bibfnamefont {M.}~\bibnamefont
  {Mukherjee}}, \bibinfo {author} {\bibfnamefont {R.}~\bibnamefont {Fazio}},
  \bibinfo {author} {\bibfnamefont {V.}~\bibnamefont {Vedral}}, \bibinfo
  {author} {\bibfnamefont {S.}~\bibnamefont {Vinjanampathy}},\ and\ \bibinfo
  {author} {\bibfnamefont {L.}~\bibnamefont {Kwek}},\ }\bibfield  {title}
  {\bibinfo {title} {Squeezing enhances quantum synchronization},\ }\href@noop
  {} {\bibfield  {journal} {\bibinfo  {journal} {Phys. Rev. Lett.}\ }\textbf
  {\bibinfo {volume} {120}},\ \bibinfo {pages} {163601} (\bibinfo {year}
  {2018})}\BibitemShut {NoStop}%
\bibitem [{\citenamefont {Laskar}\ \emph {et~al.}(2020)\citenamefont {Laskar},
  \citenamefont {Adhikary}, \citenamefont {Mondal}, \citenamefont {Katiyar},
  \citenamefont {Vinjanampathy},\ and\ \citenamefont {Ghosh}}]{expt1}%
  \BibitemOpen
  \bibfield  {author} {\bibinfo {author} {\bibfnamefont {A.~W.}\ \bibnamefont
  {Laskar}}, \bibinfo {author} {\bibfnamefont {P.}~\bibnamefont {Adhikary}},
  \bibinfo {author} {\bibfnamefont {S.}~\bibnamefont {Mondal}}, \bibinfo
  {author} {\bibfnamefont {P.}~\bibnamefont {Katiyar}}, \bibinfo {author}
  {\bibfnamefont {S.}~\bibnamefont {Vinjanampathy}},\ and\ \bibinfo {author}
  {\bibfnamefont {S.}~\bibnamefont {Ghosh}},\ }\bibfield  {title} {\bibinfo
  {title} {Observation of quantum phase synchronization in spin-1 atoms},\
  }\href@noop {} {\bibfield  {journal} {\bibinfo  {journal} {Phys. Rev. Lett.}\
  }\textbf {\bibinfo {volume} {125}},\ \bibinfo {pages} {013601} (\bibinfo
  {year} {2020})}\BibitemShut {NoStop}%
\bibitem [{\citenamefont {Koppenh{\"{o}}fer}\ \emph {et~al.}(2020)\citenamefont
  {Koppenh{\"{o}}fer}, \citenamefont {Bruder},\ and\ \citenamefont
  {Roulet}}]{expt2}%
  \BibitemOpen
  \bibfield  {author} {\bibinfo {author} {\bibfnamefont {M.}~\bibnamefont
  {Koppenh{\"{o}}fer}}, \bibinfo {author} {\bibfnamefont {C.}~\bibnamefont
  {Bruder}},\ and\ \bibinfo {author} {\bibfnamefont {A.}~\bibnamefont
  {Roulet}},\ }\bibfield  {title} {\bibinfo {title} {Quantum synchronization on
  the {IBM Q} system},\ }\href@noop {} {\bibfield  {journal} {\bibinfo
  {journal} {Phys. Rev. Research}\ }\textbf {\bibinfo {volume} {2}},\ \bibinfo
  {pages} {023026} (\bibinfo {year} {2020})}\BibitemShut {NoStop}%
\bibitem [{\citenamefont {Bastidas}\ \emph {et~al.}(2015)\citenamefont
  {Bastidas}, \citenamefont {Omelchenko}, \citenamefont {Zakharova},
  \citenamefont {Sch{\"{o}}ll},\ and\ \citenamefont {Brandes}}]{schoell_qm}%
  \BibitemOpen
  \bibfield  {author} {\bibinfo {author} {\bibfnamefont {V.~M.}\ \bibnamefont
  {Bastidas}}, \bibinfo {author} {\bibfnamefont {I.}~\bibnamefont
  {Omelchenko}}, \bibinfo {author} {\bibfnamefont {A.}~\bibnamefont
  {Zakharova}}, \bibinfo {author} {\bibfnamefont {E.}~\bibnamefont
  {Sch{\"{o}}ll}},\ and\ \bibinfo {author} {\bibfnamefont {T.}~\bibnamefont
  {Brandes}},\ }\bibfield  {title} {\bibinfo {title} {Quantum signatures of
  chimera states},\ }\href@noop {} {\bibfield  {journal} {\bibinfo  {journal}
  {Phys. Rev. E}\ }\textbf {\bibinfo {volume} {92}},\ \bibinfo {pages} {062924}
  (\bibinfo {year} {2015})}\BibitemShut {NoStop}%
\bibitem [{\citenamefont {Ishibashi}\ and\ \citenamefont
  {Kanamoto}(2017)}]{qad1}%
  \BibitemOpen
  \bibfield  {author} {\bibinfo {author} {\bibfnamefont {K.}~\bibnamefont
  {Ishibashi}}\ and\ \bibinfo {author} {\bibfnamefont {R.}~\bibnamefont
  {Kanamoto}},\ }\bibfield  {title} {\bibinfo {title} {Oscillation collapse in
  coupled quantum van der pol oscillators},\ }\href@noop {} {\bibfield
  {journal} {\bibinfo  {journal} {Phys. Rev. E}\ }\textbf {\bibinfo {volume}
  {96}},\ \bibinfo {pages} {052210} (\bibinfo {year} {2017})}\BibitemShut
  {NoStop}%
\bibitem [{\citenamefont {Amitai}\ \emph {et~al.}(2018)\citenamefont {Amitai},
  \citenamefont {Koppenh{\"{o}}fer}, \citenamefont {L{\"{o}}rch},\ and\
  \citenamefont {Bruder}}]{qad2}%
  \BibitemOpen
  \bibfield  {author} {\bibinfo {author} {\bibfnamefont {E.}~\bibnamefont
  {Amitai}}, \bibinfo {author} {\bibfnamefont {M.}~\bibnamefont
  {Koppenh{\"{o}}fer}}, \bibinfo {author} {\bibfnamefont {N.}~\bibnamefont
  {L{\"{o}}rch}},\ and\ \bibinfo {author} {\bibfnamefont {C.}~\bibnamefont
  {Bruder}},\ }\bibfield  {title} {\bibinfo {title} {Quantum effects in
  amplitude death of coupled anharmonic self-oscillators},\ }\href@noop {}
  {\bibfield  {journal} {\bibinfo  {journal} {Phys. Rev. E}\ }\textbf {\bibinfo
  {volume} {97}},\ \bibinfo {pages} {052203} (\bibinfo {year}
  {2018})}\BibitemShut {NoStop}%
\bibitem [{\citenamefont {Bandyopadhyay}\ \emph {et~al.}(2020)\citenamefont
  {Bandyopadhyay}, \citenamefont {Khatun}, \citenamefont {Biswas},\ and\
  \citenamefont {Banerjee}}]{qmod}%
  \BibitemOpen
  \bibfield  {author} {\bibinfo {author} {\bibfnamefont {B.}~\bibnamefont
  {Bandyopadhyay}}, \bibinfo {author} {\bibfnamefont {T.}~\bibnamefont
  {Khatun}}, \bibinfo {author} {\bibfnamefont {D.}~\bibnamefont {Biswas}},\
  and\ \bibinfo {author} {\bibfnamefont {T.}~\bibnamefont {Banerjee}},\
  }\bibfield  {title} {\bibinfo {title} {Quantum manifestations of homogeneous
  and inhomogeneous oscillation suppression states},\ }\href@noop {} {\bibfield
   {journal} {\bibinfo  {journal} {Phys. Rev. E}\ }\textbf {\bibinfo {volume}
  {102}},\ \bibinfo {pages} {062205} (\bibinfo {year} {2020})}\BibitemShut
  {NoStop}%
\bibitem [{\citenamefont {Bandyopadhyay}\ and\ \citenamefont
  {Banerjee}(2021)}]{qrev}%
  \BibitemOpen
  \bibfield  {author} {\bibinfo {author} {\bibfnamefont {B.}~\bibnamefont
  {Bandyopadhyay}}\ and\ \bibinfo {author} {\bibfnamefont {T.}~\bibnamefont
  {Banerjee}},\ }\bibfield  {title} {\bibinfo {title} {Revival of oscillation
  and symmetry breaking in coupled quantum oscillators},\ }\href@noop {}
  {\bibfield  {journal} {\bibinfo  {journal} {Chaos}\ }\textbf {\bibinfo
  {volume} {31}},\ \bibinfo {pages} {063109} (\bibinfo {year}
  {2021})}\BibitemShut {NoStop}%
\bibitem [{\citenamefont {Droenner}\ \emph {et~al.}(2019)\citenamefont
  {Droenner}, \citenamefont {Naumann}, \citenamefont {Sch{\"{o}}ll},
  \citenamefont {Knorr},\ and\ \citenamefont {Carmele}}]{schoellqm2}%
  \BibitemOpen
  \bibfield  {author} {\bibinfo {author} {\bibfnamefont {L.}~\bibnamefont
  {Droenner}}, \bibinfo {author} {\bibfnamefont {N.~L.}\ \bibnamefont
  {Naumann}}, \bibinfo {author} {\bibfnamefont {E.}~\bibnamefont
  {Sch{\"{o}}ll}}, \bibinfo {author} {\bibfnamefont {A.}~\bibnamefont
  {Knorr}},\ and\ \bibinfo {author} {\bibfnamefont {A.}~\bibnamefont
  {Carmele}},\ }\bibfield  {title} {\bibinfo {title} {Quantum pyragas control:
  Selective control of individual photon probabilities},\ }\href@noop {}
  {\bibfield  {journal} {\bibinfo  {journal} {Phys. Rev. A}\ }\textbf {\bibinfo
  {volume} {99}},\ \bibinfo {pages} {023840} (\bibinfo {year}
  {2019})}\BibitemShut {NoStop}%
\bibitem [{\citenamefont {Kato}\ and\ \citenamefont {Nakao}(2021)}]{qmcores}%
  \BibitemOpen
  \bibfield  {author} {\bibinfo {author} {\bibfnamefont {Y.}~\bibnamefont
  {Kato}}\ and\ \bibinfo {author} {\bibfnamefont {H.}~\bibnamefont {Nakao}},\
  }\bibfield  {title} {\bibinfo {title} {Quantum coherence resonance},\
  }\href@noop {} {\bibfield  {journal} {\bibinfo  {journal} {New J. Phys.}\
  }\textbf {\bibinfo {volume} {23}},\ \bibinfo {pages} {043018} (\bibinfo
  {year} {2021})}\BibitemShut {NoStop}%
\bibitem [{\citenamefont {Chia}\ \emph {et~al.}(2020)\citenamefont {Chia},
  \citenamefont {Kwek},\ and\ \citenamefont {Noh}}]{chia}%
  \BibitemOpen
  \bibfield  {author} {\bibinfo {author} {\bibfnamefont {A.}~\bibnamefont
  {Chia}}, \bibinfo {author} {\bibfnamefont {L.~C.}\ \bibnamefont {Kwek}},\
  and\ \bibinfo {author} {\bibfnamefont {C.}~\bibnamefont {Noh}},\ }\bibfield
  {title} {\bibinfo {title} {Relaxation oscillations and frequency entrainment
  in quantum mechanics},\ }\href@noop {} {\bibfield  {journal} {\bibinfo
  {journal} {Phys. Rev. E}\ }\textbf {\bibinfo {volume} {102}},\ \bibinfo
  {pages} {042213} (\bibinfo {year} {2020})}\BibitemShut {NoStop}%
\bibitem [{\citenamefont {Ben~Arosh}\ \emph {et~al.}(2021)\citenamefont
  {Ben~Arosh}, \citenamefont {Cross},\ and\ \citenamefont {Lifshitz}}]{lif}%
  \BibitemOpen
  \bibfield  {author} {\bibinfo {author} {\bibfnamefont {L.}~\bibnamefont
  {Ben~Arosh}}, \bibinfo {author} {\bibfnamefont {M.~C.}\ \bibnamefont
  {Cross}},\ and\ \bibinfo {author} {\bibfnamefont {R.}~\bibnamefont
  {Lifshitz}},\ }\bibfield  {title} {\bibinfo {title} {Quantum limit cycles and
  the rayleigh and van der pol oscillators},\ }\href@noop {} {\bibfield
  {journal} {\bibinfo  {journal} {Phys. Rev. Research}\ }\textbf {\bibinfo
  {volume} {3}},\ \bibinfo {pages} {013130} (\bibinfo {year}
  {2021})}\BibitemShut {NoStop}%
\bibitem [{\citenamefont {Karnatak}\ \emph {et~al.}(2007)\citenamefont
  {Karnatak}, \citenamefont {Ramaswamy},\ and\ \citenamefont
  {Prasad}}]{karnatak}%
  \BibitemOpen
  \bibfield  {author} {\bibinfo {author} {\bibfnamefont {R.}~\bibnamefont
  {Karnatak}}, \bibinfo {author} {\bibfnamefont {R.}~\bibnamefont
  {Ramaswamy}},\ and\ \bibinfo {author} {\bibfnamefont {A.}~\bibnamefont
  {Prasad}},\ }\bibfield  {title} {\bibinfo {title} {Amplitude death in the
  absence of time delays in identical coupled oscillators},\ }\href@noop {}
  {\bibfield  {journal} {\bibinfo  {journal} {Phys. Rev. E}\ }\textbf {\bibinfo
  {volume} {76}},\ \bibinfo {pages} {035201} (\bibinfo {year}
  {2007})}\BibitemShut {NoStop}%
\bibitem [{\citenamefont {Kim}\ \emph {et~al.}(2005)\citenamefont {Kim},
  \citenamefont {Roy}, \citenamefont {Aron}, \citenamefont {Carr},\ and\
  \citenamefont {Schwartz}}]{rajconj}%
  \BibitemOpen
  \bibfield  {author} {\bibinfo {author} {\bibfnamefont {M.-Y.}\ \bibnamefont
  {Kim}}, \bibinfo {author} {\bibfnamefont {R.}~\bibnamefont {Roy}}, \bibinfo
  {author} {\bibfnamefont {J.~L.}\ \bibnamefont {Aron}}, \bibinfo {author}
  {\bibfnamefont {T.~W.}\ \bibnamefont {Carr}},\ and\ \bibinfo {author}
  {\bibfnamefont {I.~B.}\ \bibnamefont {Schwartz}},\ }\bibfield  {title}
  {\bibinfo {title} {Scaling behavior of laser population dynamics with
  time-delayed coupling: Theory and experiment},\ }\href@noop {} {\bibfield
  {journal} {\bibinfo  {journal} {Phys. Rev. Lett.}\ }\textbf {\bibinfo
  {volume} {94}},\ \bibinfo {pages} {088101} (\bibinfo {year}
  {2005})}\BibitemShut {NoStop}%
\bibitem [{\citenamefont {van~der Pol}(1922)}]{vdposc}%
  \BibitemOpen
  \bibfield  {author} {\bibinfo {author} {\bibfnamefont {B.}~\bibnamefont
  {van~der Pol}},\ }\bibfield  {title} {\bibinfo {title} {On oscillation
  hysteresis in a triode generator with two degrees of freedom},\ }\href@noop
  {} {\bibfield  {journal} {\bibinfo  {journal} {Philos. Mag.}\ }\textbf
  {\bibinfo {volume} {43}},\ \bibinfo {pages} {700} (\bibinfo {year}
  {1922})}\BibitemShut {NoStop}%
\bibitem [{\citenamefont {Ermentrout}(2002)}]{xpp}%
  \BibitemOpen
  \bibfield  {author} {\bibinfo {author} {\bibfnamefont {B.}~\bibnamefont
  {Ermentrout}},\ }\href@noop {} {\emph {\bibinfo {title} {Simulating,
  Analyzing, and Animating Dynamical Systems: A Guide to Xppaut for Researchers
  and Students (Software, Environments, Tools)}}}\ (\bibinfo  {publisher} {SIAM
  Press},\ \bibinfo {year} {2002})\BibitemShut {NoStop}%
\bibitem [{\citenamefont {Johansson}\ \emph {et~al.}(2013)\citenamefont
  {Johansson}, \citenamefont {Nation},\ and\ \citenamefont {Nori}}]{qutip}%
  \BibitemOpen
  \bibfield  {author} {\bibinfo {author} {\bibfnamefont {J.}~\bibnamefont
  {Johansson}}, \bibinfo {author} {\bibfnamefont {P.}~\bibnamefont {Nation}},\
  and\ \bibinfo {author} {\bibfnamefont {F.}~\bibnamefont {Nori}},\ }\bibfield
  {title} {\bibinfo {title} {Qutip 2: A python framework for the dynamics of
  open quantum systems},\ }\href@noop {} {\bibfield  {journal} {\bibinfo
  {journal} {Comput. Phys. Commun.}\ }\textbf {\bibinfo {volume} {184}},\
  \bibinfo {pages} {1234} (\bibinfo {year} {2013})}\BibitemShut {NoStop}%
\bibitem [{\citenamefont {Gerry}\ and\ \citenamefont
  {Knight}(2005)}]{knightbook}%
  \BibitemOpen
  \bibfield  {author} {\bibinfo {author} {\bibfnamefont {C.}~\bibnamefont
  {Gerry}}\ and\ \bibinfo {author} {\bibfnamefont {P.}~\bibnamefont {Knight}},\
  }\href@noop {} {\emph {\bibinfo {title} {Introductory Quantum Optics}}}\
  (\bibinfo  {publisher} {Cambridge University Press, Cambridge, England},\
  \bibinfo {year} {2005})\BibitemShut {NoStop}%
\bibitem [{\citenamefont {Weinbub}\ and\ \citenamefont {Ferry}(2018)}]{wigner}%
  \BibitemOpen
  \bibfield  {author} {\bibinfo {author} {\bibfnamefont {J.}~\bibnamefont
  {Weinbub}}\ and\ \bibinfo {author} {\bibfnamefont {D.~K.}\ \bibnamefont
  {Ferry}},\ }\bibfield  {title} {\bibinfo {title} {Recent advances in wigner
  function approaches},\ }\href@noop {} {\bibfield  {journal} {\bibinfo
  {journal} {Appl. Phys. Rev.}\ }\textbf {\bibinfo {volume} {5}},\ \bibinfo
  {pages} {041104} (\bibinfo {year} {2018})}\BibitemShut {NoStop}%
\bibitem [{\citenamefont {Carmichael}(1999)}]{carmichael}%
  \BibitemOpen
  \bibfield  {author} {\bibinfo {author} {\bibfnamefont {H.~J.}\ \bibnamefont
  {Carmichael}},\ }\href@noop {} {\emph {\bibinfo {title} {Statistical Methods
  in Quantum Optics 1}}}\ (\bibinfo  {publisher} {Springer},\ \bibinfo {year}
  {1999})\BibitemShut {NoStop}%
\bibitem [{\citenamefont {Ansmann}(2018)}]{jitcode}%
  \BibitemOpen
  \bibfield  {author} {\bibinfo {author} {\bibfnamefont {G.}~\bibnamefont
  {Ansmann}},\ }\bibfield  {title} {\bibinfo {title} {Efficiently and easily
  integrating differential equations with {JiTCODE}, {JiTCDDE}, and
  {JiTCSDE}},\ }\href@noop {} {\bibfield  {journal} {\bibinfo  {journal}
  {Chaos}\ }\textbf {\bibinfo {volume} {28}},\ \bibinfo {pages} {043116}
  (\bibinfo {year} {2018})}\BibitemShut {NoStop}%
\bibitem [{\citenamefont {Hush}\ \emph {et~al.}(2015)\citenamefont {Hush},
  \citenamefont {Li}, \citenamefont {Genway}, \citenamefont {Lesanovsky},\ and\
  \citenamefont {Armour}}]{expt-ion}%
  \BibitemOpen
  \bibfield  {author} {\bibinfo {author} {\bibfnamefont {M.~R.}\ \bibnamefont
  {Hush}}, \bibinfo {author} {\bibfnamefont {W.}~\bibnamefont {Li}}, \bibinfo
  {author} {\bibfnamefont {S.}~\bibnamefont {Genway}}, \bibinfo {author}
  {\bibfnamefont {I.}~\bibnamefont {Lesanovsky}},\ and\ \bibinfo {author}
  {\bibfnamefont {A.~D.}\ \bibnamefont {Armour}},\ }\bibfield  {title}
  {\bibinfo {title} {Spin correlations as a probe of quantum synchronization in
  trapped-ion phonon lasers},\ }\href@noop {} {\bibfield  {journal} {\bibinfo
  {journal} {Phy. Rev. A}\ }\textbf {\bibinfo {volume} {91}},\ \bibinfo {pages}
  {061401(R)} (\bibinfo {year} {2015})}\BibitemShut {NoStop}%
\bibitem [{\citenamefont {Jayich}\ \emph {et~al.}(2008)\citenamefont {Jayich},
  \citenamefont {Sankey}, \citenamefont {Zwickl}, \citenamefont {Yang},
  \citenamefont {Thompson}, \citenamefont {Girvin}, \citenamefont {Clerk},
  \citenamefont {Marquardt},\ and\ \citenamefont {Harris}}]{expt-mem}%
  \BibitemOpen
  \bibfield  {author} {\bibinfo {author} {\bibfnamefont {A.}~\bibnamefont
  {Jayich}}, \bibinfo {author} {\bibfnamefont {J.}~\bibnamefont {Sankey}},
  \bibinfo {author} {\bibfnamefont {B.}~\bibnamefont {Zwickl}}, \bibinfo
  {author} {\bibfnamefont {C.}~\bibnamefont {Yang}}, \bibinfo {author}
  {\bibfnamefont {J.}~\bibnamefont {Thompson}}, \bibinfo {author}
  {\bibfnamefont {S.}~\bibnamefont {Girvin}}, \bibinfo {author} {\bibfnamefont
  {A.}~\bibnamefont {Clerk}}, \bibinfo {author} {\bibfnamefont
  {F.}~\bibnamefont {Marquardt}},\ and\ \bibinfo {author} {\bibfnamefont
  {J.}~\bibnamefont {Harris}},\ }\bibfield  {title} {\bibinfo {title}
  {Dispersive optomechanics: a membrane inside a cavity},\ }\href@noop {}
  {\bibfield  {journal} {\bibinfo  {journal} {New J. Phys.}\ }\textbf {\bibinfo
  {volume} {8}},\ \bibinfo {pages} {095008} (\bibinfo {year}
  {2008})}\BibitemShut {NoStop}%
\bibitem [{\citenamefont {Zakharova}\ \emph {et~al.}(2014)\citenamefont
  {Zakharova}, \citenamefont {Kapeller},\ and\ \citenamefont
  {Sch{\"{o}}ll}}]{scholl-CD}%
  \BibitemOpen
  \bibfield  {author} {\bibinfo {author} {\bibfnamefont {A.}~\bibnamefont
  {Zakharova}}, \bibinfo {author} {\bibfnamefont {M.}~\bibnamefont
  {Kapeller}},\ and\ \bibinfo {author} {\bibfnamefont {E.}~\bibnamefont
  {Sch{\"{o}}ll}},\ }\bibfield  {title} {\bibinfo {title} {Chimera death:
  Symmetry breaking in dynamical networks},\ }\href@noop {} {\bibfield
  {journal} {\bibinfo  {journal} {Phy. Rev. Lett}\ }\textbf {\bibinfo {volume}
  {112}},\ \bibinfo {pages} {154101} (\bibinfo {year} {2014})}\BibitemShut
  {NoStop}%
\bibitem [{\citenamefont {Zakharova}(2020)}]{annabook}%
  \BibitemOpen
  \bibfield  {author} {\bibinfo {author} {\bibfnamefont {A.}~\bibnamefont
  {Zakharova}},\ }\href@noop {} {\emph {\bibinfo {title} {Chimera Patterns in
  Networks}}}\ (\bibinfo  {publisher} {Springer},\ \bibinfo {address} {Cham},\
  \bibinfo {year} {2020})\BibitemShut {NoStop}%
\bibitem [{\citenamefont {Schöll}\ \emph {et~al.}(2019)\citenamefont
  {Schöll}, \citenamefont {Zakharova},\ and\ \citenamefont
  {Andrzejak}}]{sl-front}%
  \BibitemOpen
  \bibfield  {author} {\bibinfo {author} {\bibfnamefont {E.}~\bibnamefont
  {Schöll}}, \bibinfo {author} {\bibfnamefont {A.}~\bibnamefont {Zakharova}},\
  and\ \bibinfo {author} {\bibfnamefont {R.~G.}\ \bibnamefont {Andrzejak}},\
  }\bibfield  {title} {\bibinfo {title} {Editorial: Chimera states in complex
  networks},\ }\href@noop {} {\bibfield  {journal} {\bibinfo  {journal}
  {Frontiers in Applied Mathematics and Statistics}\ }\textbf {\bibinfo
  {volume} {5}},\ \bibinfo {pages} {62} (\bibinfo {year} {2019})}\BibitemShut
  {NoStop}%
\bibitem [{\citenamefont {Banerjee}(2015)}]{tanCD}%
  \BibitemOpen
  \bibfield  {author} {\bibinfo {author} {\bibfnamefont {T.}~\bibnamefont
  {Banerjee}},\ }\bibfield  {title} {\bibinfo {title}
  {Mean-field-diffusion--induced chimera death state},\ }\href@noop {}
  {\bibfield  {journal} {\bibinfo  {journal} {Europhys. Lett.}\ }\textbf
  {\bibinfo {volume} {110}},\ \bibinfo {pages} {60003} (\bibinfo {year}
  {2015})}\BibitemShut {NoStop}%
\end{thebibliography}
\providecommand{\noopsort}[1]{}\providecommand{\singleletter}[1]{#1}%
\end{document}